\documentclass[prl,twocolumn,floatfix,superscriptaddress]{revtex4-1}
\usepackage{dcolumn,amsmath}
\usepackage{graphicx}
\usepackage{bm}
\usepackage{hyperref}

\newcommand{\Eeff}{\ensuremath{E_{\rm eff}}}
\newcommand{\eEDM}{{\em e}EDM}

\newcommand{\Eref}[1]{Eq.~(\ref{#1})}

\begin{document}
\title{TaN molecule as a candidate to search for New physics}

\author{L.V.\ Skripnikov}\email{leonidos239@gmail.com}
\author{A.N.\ Petrov}
\author{N.S.\ Mosyagin}
\author{A.V.\ Titov}
\homepage{http://www.qchem.pnpi.spb.ru}

\affiliation{National Research Centre ``Kurchatov Institute'' B.P. Konstantinov Petersburg Nuclear Physics Institute, Gatchina, Leningrad district 188300, Russia}
\affiliation{Dept.\ of Physics, Saint Petersburg State University, Saint Petersburg, Petrodvoretz 198504, Russia}

\author{V. V. Flambaum}
\affiliation{School of Physics, The University of New South Wales, Sydney
NSW 2052, Australia}

\date{\today}

\begin{abstract}
It is demonstrated that the TaN molecule is the best candidate to search for T,P-violating nuclear magnetic quadrupole moment (MQM), it also looks  promising to search for other T,P-odd effects.
We report results of coupled-cluster calculations of T,P-odd effects in TaN produced by  the Ta nucleus MQM, electron electric dipole moment (EDM), scalar$-$pseudoscalar nucleus$-$electron interactions, also of  the molecule-axis  hyperfine structure constant and dipole moment. 
Nuclear calculations of $^{181}$Ta MQM are performed to express the T,P-odd effect in terms of the strength constants of T,P-odd nuclear forces, proton and neutron EDM, QCD parameter $\theta$ and quark chromo-EDM.
\end{abstract}
\pacs{
31.30.jp,
32.10.Fn,
31.15.vn
 }

\maketitle

\section{Introduction}

During last years low-energy experiments on heavy atoms and diatomic molecules containing heavy atoms have proved to be very important for the  search for  New physics beyond the standard model \cite{Regan:02, Hudson:11a, ThO}.
The best limit on the electron electric dipole moment (\eEDM) was obtained on the ThO molecular beam in \cite{ThO}. The experiment was also sensitive to other effect that violates time-reversal (T) and parity symmetries (P) -- scalar$-$pseudoscalar nucleus$-$electron neutral current interactions \cite{Pospelov:14}. If a heavy atom has nuclear spin I$>1/2$ one can expect that the molecule will also be sensitive to other T,P-odd effect, the T,P-odd interaction of nuclear magnetic quadrupole moment (MQM) with electrons 
\cite{FKS84b,KhMQM,F94}. Important ideas on the subject were suggested in papers  \cite{Sandars:64, Sandars:67, Labzowsky:78, Sushkov:78, FKS84b,KhMQM,F94}.  
%see also reviews \cite{GFreview, Khriplovich:11} and references therein.

There is a number of systems on which experiments to search for T,P-odd effects have already been conducted or suggested and which are investigated theoretically and experimentally (HfF$^+$ \cite{Cossel:12, Cornell:13, Petrov:07a, Fleig:13, Meyer:06a,Skripnikov:08a,Le:13}, YbF \cite{Hudson:11a, Mosyagin:98, Quiney:98, Parpia:98, Kozlov:97c, Nayak:09, Steimle:07, Abe:14}, ThO \cite{ThO, Petrov:14, Meyer:08, Skripnikov:13c, Skripnikov:14a, Skripnikov:14b, Fleig:14}, ThF$^+$~\cite{Cornell:13,Skripnikov:15b}, WC \cite{Lee:13a, Meyer:09a}, PbF \cite{Shafer-Ray:08E, Skripnikov:14c, Petrov:13,Skripnikov:15d}, RaO \cite{Flambaum:08,Kudashov:13}, RaF \cite{Isaev:12,Kudashov:14},  PtH$^+$ \cite{Meyer:06a, Skripnikov:09}, etc.).

Recently, TaN molecule was suggested as a new system to search for the T,P-odd MQM of the tantalum nucleus \cite{FDK14}, where the molecule was marked as quite promising candidate to search for T,P violation in nuclear sector using molecules.
 Indeed, $^{181}$Ta nucleus is stable and its MQM is strongly enhanced due to the collective effect \cite{F94}, large nuclear charge $Z$ of Ta leads to a higher than $Z^2$ electron enhancement of the T,P-odd  effects \cite{Sandars:67,Sushkov:78,KhMQM,FKS84b}, TaN in $^3\Delta_1$ state has $\Omega$-doublets which allow full polarization in a small electric field and  cancellation of many systematic errors due to opposite signs of the effects on the doublet components \cite{Sushkov:78,DeMille:2001}, very small magnetic moment due to cancellation of the orbital and spin contributions \cite{Meyer:06a} and long lifetime of    $^3\Delta_1$ electron state  \cite{FDK14}.

TaN molecule has the triple bond order, i.e.\ higher than in some other recently experimentally considered systems to search for T,P-violation effects, e.g.\ ThO molecule \cite{ThO}. This suggests a rather complicated electronic structure in terms of electron correlation effects. Up to now the molecule was rather poor investigated both experimentally and theoretically \cite{Ram:02,FDK14}. There are no experimental data about the molecule-frame dipole moment, hyperfine structure, etc. Therefore, the aim of the present paper is to perform the first reliable {\it ab~initio} treatment of the TaN electronic structure in $^3\Delta_1$ state including different T,P-violating effects. Also we provide the values of the molecule-frame dipole moment and molecule-axis hyperfine structure constant which can be important for experimental planning and determination of the quantities and acceleration of the further experimental treatment of the system. To solve the problem we have developed pure {\it ab~initio} scheme of the calculation where electron correlation effects were considered within the most accurate all-order coupled-cluster method, in which we took account of up to quadruple cluster amplitudes.

%\section{Theory}

Spin-dependent T,P-odd Hamitonian can be expressed in the following form \cite{Kozlov:87, Dmitriev:92}:
\begin{gather}
H^{\rm T,P} = 
    H^{\rm SP} +  H^{d} + H^{\rm MQM}
\label{HEFF}
\end{gather}
The T,P-odd scalar$-$pseudoscalar nucleus$-$electron Hamiltonian $H^{\rm SP}$ with a characteristic dimensionless constant $k_{\rm SP}$ is given by \cite{Hunter:91}:
\begin{eqnarray}
   H^{\rm SP}=i\frac{G_F}{\sqrt{2}}Z k_{\rm SP}\gamma_0\gamma_5
\rho_N(\textbf{r}),
 \label{HS}
\end{eqnarray}
where $Z$ is the heavy nucleus charge, $G_F$ is the Fermi-coupling constant, $\gamma_0$ and $\gamma_5$ are the Dirac matrices and $\rho_N(\textbf{r})$ is the nuclear density normalized to unity.
For the interaction of~ \eEDM~ with inner molecular electric field, $\bm{E}$, one has
\begin{eqnarray}
  H^d=2d_e
  \left(\begin{array}{cc}
  0 & 0 \\
  0 & \bm{\sigma E} \\
  \end{array}\right)\ ,
 \label{Hd}
\end{eqnarray}
where $d_e$ is the value of \eEDM, $\bm{\sigma}$ are the Pauli matrices.
Hamiltonian of interaction of MQM with electrons is given by \cite{Kozlov:87,GFreview}
%\footnote{The T,P-violating magnetic quadrupole moment (\ref{eqaux1}) gives rise to a vector potential, $\vec{A}^{\rm MQM}$, see Eqs. (165-167) in Ref.~\cite{GFreview}. Substituting $\vec{A}^{\rm MQM}$ to the Dirac equation we go to the interaction $|e| (\vec{\alpha} \cdot \vec{A}^{\rm MQM})$ coinciding with Eq.~(\ref{hamq}).}:
%
%
 \begin{align}\label{hamq}
% H  &=
 H^{\rm MQM}  &=
 -\frac{  M}{2I(2I-1)}  T_{ik}\frac{3}{2} \frac{[\bm{\alpha}\times\bm{r}]_i r_k}{r^5},
 \end{align}
where Einstein's summation convention is implied, $\bm{\alpha}$ are the 4x4 Dirac matrices,
$ \bm{\alpha}=
  \left(\begin{array}{cc}
  0 & \bm{\sigma} \\
  \bm{\sigma} & 0 \\
  \end{array}\right),
$
 $\bm{r}$ is the displacement of the electron from the Ta nucleus, $\bm I$ is the nuclear spin,
  $M$ is the nuclear MQM,
\begin{align}\label{eqaux1}
M_{i,k}=\frac{3M}{2I(2I-1)}T_{i,k}\, \\
 T_{i,k}=I_i I_k + I_k I_i -\tfrac23 \delta_{i,k} I(I+1)\,.
 \end{align} 
In the subspace of $\pm \Omega$ states ($\Omega= \langle\Psi|\bm{J}\cdot\bm{n}|\Psi\rangle$,
$\bm{J}$ is the total electronic momentum, $\Psi$ is the {\it electronic} wave function for the considered $^3\Delta_1$ state of TaN) the expressions (\ref{HS},\ref{Hd},\ref{hamq}) are reduced to the following effective molecular Hamiltonians, correspondingly~\cite{FKS84b}:
\begin{equation}
H^{\rm SP}_\mathrm{eff} = W_{\rm SP}~k_{\rm SP} \mathbf{S}^{\prime}\cdot\mathbf{n},
\end{equation}
\begin{equation}
H^{d}_\mathrm{eff} = W_{d}~d_{e}\mathbf{S}^{\prime}\cdot\mathbf{n},
\end{equation}
 \begin{align}\label{eq0}
% H_\mathrm{eff} &=
H^{\rm MQM}_\mathrm{eff} &=
 -\frac{W_M  M}{2I(2I-1)} \bm S^{\prime} \hat{\bm T} \mathbf{n}
 \,,
 \end{align}
where $\mathbf{n}$ is the unit vector along the molecular axis $\zeta$ directed from Ta to N, $\bm S^{\prime}$ is the effective electron spin~\cite{Kozlov:95}
defined by the following equations: 
$\mathbf{S}^{\prime}_\mathbf{\zeta}|\Omega> = \Omega|\Omega>$,
$\mathbf{S}^{\prime}_{\pm}|\Omega=\pm 1> = 0$ \cite{Kozlov:87, Dmitriev:92}, $S{=}|\Omega|{=}1$.

To extract the fundamental parameters $k_{\rm SP}$, $d_e$, $M$  from an experiment one needs to know the factors $W_{\rm SP}$, $W_d$ and $W_M$, correspondingly, which are determined by the electronic structure of a studied molecular state on a given nucleus (discussed in Refs.~\cite{Kozlov:87, Kozlov:95, Titov:06amin, FDK14}):
\begin{equation}
\label{WS}
W_{\rm SP} = \frac{1}{\Omega}
\langle \Psi|\sum_i\frac{H^{\rm SP}(i)}{k_{\rm SP}}|\Psi
\rangle\,
\end{equation}
\begin{equation}
\label{matrelem}
W_d = \frac{1}{\Omega}
\langle \Psi|\sum_i\frac
{H^d(i)}
{d_e}|\Psi\rangle,
\end{equation}
\begin{align}
  \label{WM}
W_M= 
\frac{3}{2\Omega} 
   \langle
   \Psi\vert\sum_i\left(\frac{\bm{\alpha}_i\times
\bm{r}_i}{r_i^5}\right)
 _\mathbf{\zeta} r_\mathbf{\zeta} \vert
   \Psi\rangle\ .
 \end{align}
Note, that a parameter known as the effective electric field acting of unpaired electrons, $E_{\rm eff}=W_d|\Omega|$, is often used.

For a completely polarized molecule the energy shift due to \eEDM, scalar$-$pseudoscalar nucleus$-$electron neutral current, and MQM interactions are
\begin{equation}
\label{shiftd}
\delta_d = d_eW_d\Omega,
\end{equation}
\begin{equation}
\label{shiftd}
\delta_{\rm SP} = k_{\rm SP}W_{\rm SP}\Omega,
\end{equation}
\begin{eqnarray}
\label{shiftM}
%\delta(J,F) = (-1)^{\Omega+I+F+1}C(J,F) W_M M\ , \\
\delta_M(J,F) = (-1)^{I+F}C(J,F) M W_M \Omega\ , \\
C(J,F)= \frac{(2J+1)}{2}
\frac{
    \left(
    \begin{array}{ccc}
    J &  2 &  J \\
   -\Omega & 0 & \Omega
    \end{array}
    \right)
    }
    {
    \left(
    \begin{array}{ccc}
    I &  2 &  I \\
   -I & 0 & I
    \end{array}
    \right)
    }
    \left\{
    \begin{array}{ccc}
    J &  I &  F \\
    I &  J & 2
    \end{array}
    \right\},
\end{eqnarray}
where $(...)$ means elements with 3j$-$symbols and $\{...\}$ with 6j$-$symbols \cite{LL77}, $F$ is the total angular momentum and $J$ is the number of rotational level.
 Note, that both $\delta_d$ and $\delta_{\rm SP}$ are independent
of $J$ and $F$ quantum numbers, 
whereas $\delta_M$ depends on them. Also, in opposite to $H^{\rm SP}_\mathrm{eff}$ and $H^{d}_\mathrm{eff}$ Hamiltonians
 $H^{\rm MQM}_\mathrm{eff}$ has  non-zero off-diagonal matrix elements on $J$ quantum number (between different rotational levels).
This should be taken into account when mixing of different rotational levels become significant. In \Eref{shiftM} this effect is neglected.
For $^{181}$TaN ($I{=}7/2$) and ground rotational level {$J{=}1$} \Eref{shiftM} gives the MQM energy shifts, $\left| \delta(J,F) \right|$, equal to $0.107 W_M M, \ 0.143 W_M M, \ 0.05 W_M M$ for $F= 5/2,7/2,9/2$, correspondingly.

To compute hyperfine structure constant $A_{||}$ on $^{181}$Ta in the $^3\Delta_1$ electronic state of $^{181}$TaN molecule the following matrix element can be evaluated:
\begin{eqnarray}
\label{Apar}
%\nonumber
A_{\parallel} = 
  \frac{\mu_{\rm Ta}}{I\Omega}
   \langle
   \Psi|\sum_i\left(\frac{\bm{\alpha}_i
\times
\bm{r}_i}{r_i^3}\right)
_\mathbf{\zeta}
|\Psi\rangle,
\end{eqnarray}
where $\mu_{\rm Ta}$ is the nuclear magnetic moment of a Ta isotope
with spin $I$.

\section{Nuclear Magnetic Quadrupole moment}

The main contribution to MQM is produced by the nucleon-nucleon T,P-odd interaction  which exceeds  the nucleon EDM contribution  by 1-2 orders of magnitude \cite{FKS84b}. In a spherical nucleus MQM is determined mainly by a valence nucleon which carries the nuclear angular momentum $I$ \cite{FKS84b}:
\begin{align}\label{M}
%M=M_0^v (2I-1)t_I\,,\\
%\label{M0}
M=[d-2\cdot 10^{-21} \eta(\mu-q)  (e \cdot {\rm cm})] \lambda_p (2I-1)t_I,
 \end{align}
where $t_I=1$ for $I=l+1/2$ and $t_I=-I/(I+1)$ for $I=l-1/2$, $I$ and $l$ are the total and orbital angular momenta of a valence nucleon,
 $\eta$ is the dimensionless strength constant of the T,P-odd nuclear potential $\eta G/(2^{3/2} m_p) (\sigma \cdot \nabla \rho)$ acting on the valence nucleon,  $\rho$ is the total nucleon number density,
 the nucleon magnetic moments are $\mu_p=2.79$ for valence proton  and $\mu_n=-1.91$ for valence neutron, $q_p=1$ and $q_n=0$, $\lambda_p=\hbar /m_pc=2.10 \cdot 10^{-14}$ cm,
the  contribution of the valence nucleon EDM $d$ was calculated in Ref. \cite{KhMQM} . 

In deformed nuclei MQM has a collective nature and is enhanced by  an order of magnitude \cite{F94} (comparable to  the enhancement of an ordinary collective electric quadrupole moment). An estimate of the collective MQM can be done as follows.
In a deformed nucleus the strong field splits orbitals with different absolute values $|I_z|$ of the projections of the angular momentum  on the nuclear symmetry axis.
 Sum over all $I_z$ gives zero contribution to MQM. However, due to the difference of energies for different $|I_z|$ some of the $I_z$ orbitals are vacant, and the remaining contribution is not zero.   As a result, the MQM of a deformed nucleus in the ``frozen'' frame (rotating together with a nucleus) may be estimated using the following formula \cite{F94}:
 \begin{align}\label{Mzz}
M^{\rm nucl}_{zz}=\sum M^{\rm single}_{zz} (I,I_z,l) n(I,I_z,l),
 \end{align}
where the sum goes over occupied orbitals,  $M^{\rm single}_{zz}(I,I_z,l)$ is given by Eqs.~(\ref{M}) and (\ref{eqaux1}), $T_{zz}=2 I_z^2 -\tfrac23 I(I+1)$, $n(I,I_z,l)$ are the orbital occupation numbers, which may be found in Ref. \cite{Bohr}. The sum over a complete shell gives zero; therefore, for shells more than half-filled it is convenient to use hole numbers in place of particle numbers, using the relation $M^\mathrm{single}_{zz} (\mathrm{hole})=- M^\mathrm{single}_{zz}(\mathrm{particle})$.

The nucleus $^{181}$Ta  has  the following occupation numbers: 10 neutron holes in orbitals  $[\bar{l}_I, I_z] = [\bar{p}_{3/2}, \pm 3/2]$, $[\bar{i}_{13/2},\pm 13/2, \pm 11/2 ]$, $[\bar{h}_{9/2}, \pm 9/2, \pm 7/2 ]$, and 9 proton holes $[\bar{d}_{5/2}, \pm 5/2 ]$,  
 $[\bar{h}_{11/2}, \pm 11/2, \pm 9/2]$, 
$[\bar{d}_{3/2}, \pm 3/2 ]$,  $[\bar{g}_{7/2}, 7/2 ]$.

The MQM in the laboratory frame $M\equiv M_{\rm lab}$ can be expressed via MQM in the rotating frame (\ref{Mzz}):
\begin{align}\label{Mlab}
% M^{\rm lab}=\frac{J(2J-1)}{(J+1)(2J+3)} M^{\rm nucl}_{zz}=\\
\nonumber 
 M^{\rm lab}=\frac{I(2I-1)}{(I+1)(2I+3)} M^{\rm nucl}_{zz}=\\
\nonumber
(1.1 \eta_p- 0.9  \eta_n) \cdot 10^{-33}  (e \cdot {\rm cm}^2)\\
-(3.0 d_p +2.3 d_n) \cdot 10^{-13}  {\rm cm},
 \end{align}
where 
% $J=5/2$ 
  $I=7/2$ 
is the  nuclear spin of $^{181}$Ta.

The T,P-odd nuclear forces are dominated by the $\pi_0$ meson exchange \cite{FKS84b}. Therefore, we may express the strength constants via strong $\pi NN$ coupling constant $g=13.6$ and T,P-odd $\pi N N$ coupling constants corresponding to the isospin channels  $T=0,1,2$:  $\eta_n= - \eta_p = 5\cdot 10^6 g (  {\bar g_1}{+}0.4{\bar g_2}{-}0.2{\bar g_0}) $ (see detailes in \cite{Skripnikov:14a}). 
As a result, we obtain 
\begin{align}\label{Mg}
M(g)= - [g (  {\bar g_1}+ 0.4 {\bar g_2}-0.2 {\bar g_0})  \cdot  8 \cdot 10^{-27} e \cdot {\rm cm}^2 .
 \end{align}
Possible CP-violation in the strong interaction sector is described by the  CP violation parameter ${\tilde \theta}$. According to Ref.~\cite{theta}  $g {\bar g_0}=-0.37 {\tilde \theta} $. This gives the following value of MQM for $^{181}$Ta:
\begin{align}\label{Mtheta}
M(\theta) = -5 \cdot 10^{-28} {\tilde \theta} \cdot e \cdot {\rm cm}^2 .
 \end{align} 
Almost the same final results for $M(\theta)$ can be obtained by using recently calculated relations of $g_0$ and $g_1$ constants by using chiral perturbation theory \cite{Dekens:14} and lattice data for the strong part of the proton-neutron mass difference \cite{Bsaisou:15} which give updated value of $gg_0$.

Finally, we can express MQM in terms of the quark chromo-EDM ${\tilde d_u}$ and  ${\tilde d_d}$ using the relations 
 $g {\bar g_1}=4.{\cdot}10^{15}( {\tilde d_u} - {\tilde d_d})/{\rm cm} $, $g {\bar g_0}=0.8 \cdot 10^{15}( {\tilde d_u} + {\tilde d_n})/{\rm cm} $ 
 \cite{PospelovRitzreview}:
\begin{align}\label{Md}
 M( {\tilde d}) = -3 \cdot 10^{-11} ( {\tilde d_u} - {\tilde d_d}) \cdot e \cdot {\rm cm} .
 \end{align}
The contributions of $d_p$  and  $d_n$ to MQM  in Eqs.~(\ref{Mg} -\ref{Md}) are from one to two orders of magnitude smaller than the contributions of the nucleon CP-odd interactions.
%end

\section{Electronic structure calculation details}

To obtain the electronic state-specific parameters $W_{\rm SP}$, \Eeff, $W_M$ and $A_{||}$ described by Eqs.~(\ref{matrelem},\ref{WS},\ref{WM},\ref{Apar}) we have performed a series of calculation with using the two-step procedure to study the relativistic four-component electronic structure in the vicinity of the Ta nucleus \cite{Titov:06amin, Petrov:02, Skripnikov:15b}. 
For this the space around the given heavy atom, Ta, is divided into the valence and core regions. At the first step inactive core electrons are excluded from molecular calculations using the generalized relativistic effective core potential (GRECP) method \cite{Titov:99, Mosyagin:10a}.
The approach allows one also to take into account the contribution from Breit interaction and finite nuclear size \cite{Petrov:04b, Mosyagin:06amin}.
 After this stage we obtain wavefunction that is very accurate in the valence region but has incorrect behaviour in the core region. The correct four-component behaviour of the wave-function in the core region is restored at the second step using the procedure 
\cite{Titov:06amin, Skripnikov:15b}  based on a proportionality of valence and virtual (unoccupied in the reference Slater determinant) spinors \cite{Titov:14a} in the inner-core regions of heavy atom.
The splitting of solution of the four-component relativistic full-electron problem on two consequent steps allows one to consider high-order correlation effects (see below) that are important for reliable and accurate calculation of the properties that cannot be measured experimentally.

The applied computation scheme includes the following steps:
(i) Consideration of leading correlation and relativistic effects within the relativistic two-component coupled-cluster with single, double and perturbative triple cluster amplitudes.
The correlation calculations included valence and outer-core electrons of Ta ($5s^25p^66s^25d^3$) and N ($1s^22s^22p^3$), i.e.\ 20 electrons.
For Ta we used the basis set consisting of 15 $s-$, 10 $p-$, 10 $d-$, 5 $f-$ and 2 $g-$ type uncontracted Gaussians 
  \footnote{Used basis sets are available at http://qchem.pnpi.spb.ru website}. 
For N we used the aug-ccpVQZ basis set \cite{Kendall:92} with removed two g-type basis functions.
Below this basis set will be called MBas.
(ii) Consideration of correction on basis set enlargement. 
It was calculated as a difference between results of 20-electron scalar-relativistic calculation within the CCSD(T) method in enlarged basis 
\footnote{Basis for Ta was increased up to 18 $s-$, 18 $p-$, 15 $d-$, 15 $f-$, 10 $g-$, 7 $h-$ and 7 $i-$ type Gaussian basis functions.}
 and basis used at step (i).
(iii) Consideration of high-order correlation effects.
They were calculated as a difference between results of 20-electron two-component calculations within the couple cluster with single, double, triple and perturbative quadruple cluster amplitudes, the CCSDT(Q), and the CCSD(T) calculation with using a compact basis set \cite{Skripnikov:14b, Skripnikov:13a}.
(iv) Consideration of contribution of sub-outer-core $4s^2 4p^6 4d^{10} 4f^{14}$ electrons of Ta. Their contribution was obtained as a difference between the results of 52-electron and 20-electron 2-component CCSD(T) calculations \footnote{In the 20-electron calculations $4s^2 4p^6 4d^{10} 4f^{14}$ electrons were frozen from atomic calculation, i.e.\ in the spherical form rather than the relaxed molecular form.
For Ta we used basis set consisting of 18 $s-$, 18 $p-$, 10 $d-$, 7 $f-$ and 4 $g-$ type Gaussian basis functions
 }.
Core electrons of Ta ($1s-4f$) in steps (i), (ii) and (iii) were excluded from correlation treatment by the 60-electron GRECP.
% generated in the present study. 
%
For step (iv) we constructed the GRECP version for 45 explicitly treated electrons of the Ta atom and a very small 28-electron inner core. 

The coupled-cluster calculations were performed using the {\sc mrcc} \cite{MRCC2013,Kallay:6} code interfaced to {\sc dirac} code \cite{DIRAC12}. Scalar-relativistic calculations were performed using {\sc cfour} program package \cite{CFOUR}. Restoration of four-component electronic structure was performed using the code developed in \cite{Skripnikov:11a, Skripnikov:13b, Skripnikov:15b} and interfaced to the above-mentioned packages.

Equilibrium Ta$-$N distance and other spectroscopic parameters were obtained by approximation of potential energy curve calculated within the 20-electron two-component CCSD(T) method in MBas basis set.

\section{Results and discussions}
%################################

The equilibrium internuclear distance calculated for the $^3\Delta_1$ state of TaN is 3.19 a.u.\ which agrees well with the experimental datum \cite{Ram:02}, see table \ref{spec_props}. In calculations of the parameters under consideration we have set R(Ta--N) to 3.20~a.u.

\begin{table}[b]
  \caption{Equilibrium internuclear distance $R_e$, harmonic vibrational wavenumber $\omega_e$ and vibrational anharmonicity $\omega_e x_e$ for the $^3\Delta_1$ state of TaN.}
  \begin{tabular}{lccc}
   \hline \hline
   Method                & $R_e$, a.u. & $\omega_e$, cm$^{-1}$& $\omega_e x_e$, cm$^{-1}$ \\
   \hline   
   20e-2c-CCSD(T),    this work           & 3.19 & 1028 & 3.5  \\
   Experiment, \cite{Ram:02}    & 3.20 & ---  & ----	     \\   
   \hline \hline
  \end{tabular}
  \label{spec_props}
\end{table}

TaN has triple bond order (one $\sigma$-bond and two $\pi$ bonds) provided mainly by $d$-electrons of Ta and $p$-electrons of N \cite{Mayer:07,Sizova:08,Sizova:08b,Sizova:09}. Unpaired electrons in $^3\Delta_1$ state are non-bonding electrons localized on Ta. They determine the $\sigma^1\delta^1$ configuration, where $\sigma$ is mainly the $6s$ atomic orbital of Ta and $\delta$ is mainly $5d$ atomic orbital of Ta.

\begin{table*}[!h]
\caption{
The calculated values of the molecule-frame dipole moment ($d$), effective electric field (\Eeff), parameter of the T,P-odd scalar$-$pseudoscalar nucleus$-$electron interaction ($W_{\rm SP}$), parameter of T,P-odd MQM interaction ($W_M$), hyperfine structure constant (A$_{||}$) of the $^3\Delta_1$ state of TaN using the coupled-cluster methods. 
}
\label{TResultsEeff}
\begin{tabular}{l l  r  r  r  r  r  r }
\hline\hline
 Method   & $d$,  & \Eeff, & $W_{\rm SP}$, & $W_M$ & A$_{||}$ $^a$,    \\
            & Debye & GV/cm & kHz & $\frac{10^{33}\mathrm{Hz}}{e~{\rm cm}^2}$ & 
MHz  \\

\hline
  20e-2c-CCSD                 & {} 4.70 & {} 37.5 & 34  & 1.15  &  -3082   \\  
  20e-2c-CCSD(T)              & {} 4.81 & {} 35.6 & 32  & 1.10  &  -3015   \\ 
  correlation correction      & {}-0.04 & {}  0.2 &  0  &-0.01  &  -32     \\    
  basis set correction        & {}-0.03 & {} -0.4 &  0  & 0.00  &  -33     \\ 
  core correction             & {} 0.00  & {} -0.5 &  -1 &-0.01  &  -51     \\
  \textbf{FINAL}              & {} 4.74  & {} 34.9 &  31 & 1.08  &  -3132   \\  

\hline\hline
\end{tabular}

$^a$ Magnetic moment of Ta ($\mu_{^{181}Ta}$) was set to 2.371$\mu_{\rm N}$ \cite{Mills:88,Erich:73}.

\end{table*}

The calculated values of \Eeff, $W_{\rm SP}$, $W_M$ and $A_{||}$ are given in table \ref{TResultsEeff}.
The effective electric field is rather strong (1.5 times more than that in HfF$^+$~\cite{Petrov:07a, Fleig:13} and similar to other considered transition elements compounds \Eeff(PbF)\cite{Skripnikov:14c} and WC~\cite{Lee:13a}) but about two times smaller than \Eeff\ in ThO \cite{Skripnikov:13c,Skripnikov:14b, Fleig:14}. The value of the $W_M$ parameter estimated in Ref.~\cite{FDK14} ($\approx 1$) agrees with the value obtained in this paper.
Thus the calculated large value of the $W_M$ parameter confirms that the $^{181}$TaN molecule with stable $^{181}$Ta isotope having nucleus spin I$>1/2$ probably the most promising candidate to-date among considered heavy-atom diatomic molecules to search for nuclear MQM.
According to the results listed in table \ref{TResultsEeff} and our previous error analysis from Ref.~\cite{Skripnikov:14b} for a comparable situation (ThO in $^3\Delta_1$ state) we expect that the theoretical uncertainty of the calculated  characteristics is less than 7\%.
Direct experimental check of \Eeff, $W_{\rm SP}$  and $W_M$ parameters is impossible, however, one can in principle seize here the opportunity of  measurement of the hyperfine structure constant in $^3\Delta_1$ state of TaN which can be performed later. As it was argued earlier \cite{Kozlov:97, Titov:06amin, Titov:14a} the ``equal-footing calculation'' of the the hyperfine structure constants can provide a very important (though indirect) test for the \Eeff, $W_{\rm SP}$  and $W_M$ parameters.
The calculated value of the molecule-axis hyperfine structure constant is given in table \ref{TResultsEeff}.

One can express the MQM energy shift, $(-1)^{I+F}C(J,F) M W_M \Omega$ in terms of the fundamental CP-violating physical quantities $d_p$, $d_n$, $\tilde{\theta}$ and $\tilde{d}_{u,d}$ using Eqs.~(\ref{Mlab},\ref{Mtheta},\ref{Md}). For the lowest rotational level, for which the coefficient $|C(J{=}1,F{=}7/2)|=0.143$ 
reaches a maximum value, we have
\begin{align} 
\label{EprotonEDM} 
0.143 W_M  M  &=  
- \frac{10^{25}(4.5  d_p+3.5 d_n)}{e \cdot \mathrm{cm}} \cdot \mu{\rm Hz} 
\end{align} 
\begin{align}
\label{EshiftTheta}
0.143 W_M  M  &= -7.7 \cdot 10^{10} {\tilde \theta} \cdot \mu{\rm Hz} 
\end{align}  
\begin{align} 
\label{EshiftD} 
0.143 W_M  M  &= -4.6 \cdot \frac{10^{27}({\tilde d_u}-{\tilde d_d})}{\mathrm{cm}} \cdot \mu{\rm Hz}
\end{align} 
The current limits on $d_p$, $|{\tilde \theta}|$ and $|{\tilde d_u}{-}{\tilde d_d}|$ ($|d_p|< 8.6 \cdot 10^{-25} e \cdot $cm, $|{\tilde \theta}| < 2.4 \cdot 10^{-10}$, $|{\tilde d_u}{-}{\tilde d_d}|<6 \cdot 10^{-27} $~cm ~\cite{Hg} correspond to the shifts $|0.143\ W_M M | < 40~\mu$Hz, $18~\mu$Hz and $ 28 ~\mu$Hz, respectively.
Currently the best limit on the energy shift produced by the T,P-odd effects (electron EDM and scalar$-$pseudoscalar nucleus$-$electron interactions) in $^3\Delta_1$ state of $^{232}$ThO is $700~\mu$Hz \cite{ThO} and is expected to be improved by an order of magnitude over the next five years \cite{Hess:14,Spaun:14} due to experimental developments. Such a progress with the $^{181}$TaN experiment on the $^3\Delta_1$ state can result in comparable improvement with limitations on the T,P-odd effects in the nuclear sector as well.

\section*{Acknowledgement}
The generalized relativistic effective core potentials were generated with the support of RFBR Grant No.~13-03-01307. The basis sets for Ta were generated with the support of 
the Russian Science Foundation grant (project No.\ 14-31-00022).
%The molecular calculations were partly performed on the Supercomputer ``Lomonosov''.
The rest of the work was supported by the following grants.
L.S., N.M., A.T.\ and A.P.\ acknowledge support from  Saint Petersburg 
State University, research grant 0.38.652.2013 and RFBR Grant No.~13-02-0140.
L.S.\ is also grateful to the grant of President of Russian Federation 
No.MK-5877.2014.2 and Dmitry Zimin ``Dynasty'' Foundation.
V.F.\ acknowledges support from Australian Research Council and Humboldt 
Research Award. 
We thank Profs. Andreas Wirzba, Christoph Hanhart and Jordy de Vries for pointing us about recent data about $g$, $g_0$, $g_1$ constants to obtain $M(\theta)$ dependence.

%\bibliographystyle{./bib/apsrev}
%\bibliography{bib/JournAbbr,bib/SkripnikovLib,bib/QCPNPI,bib/TitovLib,bib/Kaldor,bib/PetrovLib,bib/Lomachuk,bib/Kudashov}

\end{document}